# Audio classification using ML methods

Krishna Kumar
M.Tech Artificial Intelligence
*REVA Academy for Corporate Excellence - RACE, REVA University*
Bengaluru, India
krishna.kumar1612@gmail.com

*Abstract*— Machine Learning systems have achieved outstanding performance in different domains. In this paper machine learning methods have been applied to classification task to classify music genre. The code shows how to extract features from audio files and classify them using supervised learning into 2 genres namely classical and metal. Algorithms used are LogisticRegression, SVC using different kernals (linear, sigmoid, rbf and poly), KNeighborsClassifier , RandomForestClassifier, DecisionTreeClassifier and GaussianNB.

## I. INTRODUCTION

Machine Learning is used to classify the audio files into 2 genres classical and metal. A total of 20 audio files, 10 for each genre respectively are taken. The dataset is prepared by extracting features. 138 features from each file. The dataset is of shape 20 x 138. The application of ML to classify is shown in 2 parts. 1st part shows feature engineering using LDA - Linear discriminant analysis. In part 2nd an expert opinion is taken for feature selection (here it is randomly done). The code and the audio data files are present in GitHub location https://github.com/krishnakumar1612/music_genre_classification.

## II. RELATED WORK

Python package pyAudioAnalysis is used to extract the 138 features for each audio file. This is a supervised machine learning classification problem, and we know the genre (classical, metal) of each music file. This is a balanced data set with 20 samples, 10 for each genre. The dataset is a shape of 20 x 138. It is clear there is a need to have some feature engineering or feature selection as the number of columns is huge in comparison with rows. Also, it is clear from the data that there is a need to normalize. StandardScaler is used to preserve the data distribution. Next Machine Learning models LogisticRegression KNeighborsClassifier RandomForestClassifier DecisionTreeClassifier GaussianNB and SVC with kernal 'linear' 'poly' 'rbf' and 'sigmoid' are trained. The results are discussed below in 2 parts. In part 1 we have applied feature engineering using LDA- linear discriminant analysis and in part 2 we apply feature selection with some expert opinion.

### A. Part 1 Feature Engeneering

LDA is applied to get a single feature which explains the labels most. We can get only 1 feature as LDA n_components because the number of classes are 2 namely 'classical', 'metal'.

We can see this new feature distribution to get an idea of algo we can apply.

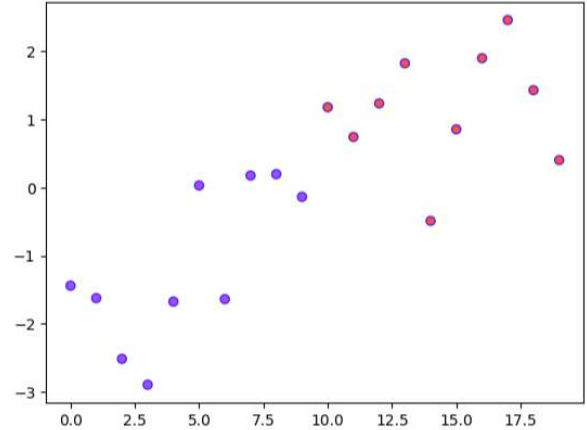

Fig. 1. Feature distribution in 2d

From figure 1 data seems to be linearly separable. Next, we applied to KNN.

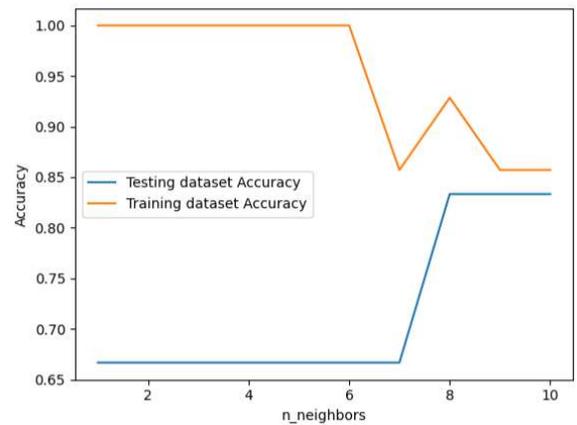

Fig. 2. Value of k vs accuracy

From fig 2 k=8 is the best choice for n_neighbors parameter. For accuracy calculation, complete dataset is used and not only test and train data separately. This is due to the very less number of samples, 20 in our data set. From confusion matrix we see that Logistic regression and SVC with kernel sigmoid has an accuracy of 0.95. Below is the confusion matrix for the KNeighborsClassifier.

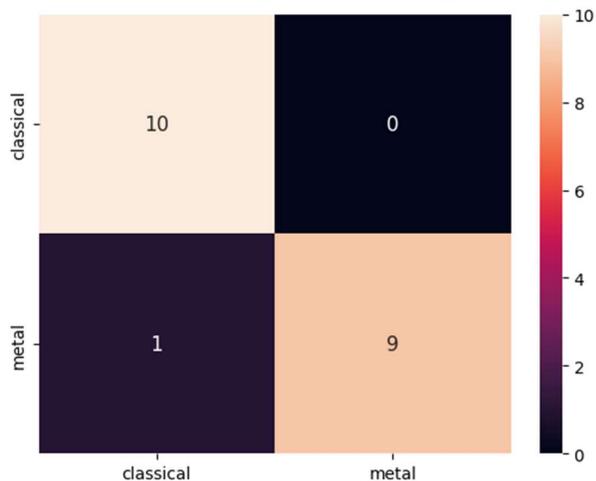

Fig. 3. Confusion matrix for KNeighbour classifier

As expected, the accuracy for SVC support vector classifier with kernel 'poly' is worst with accuracy value 0.75. Below is the confusion matrix for the same.

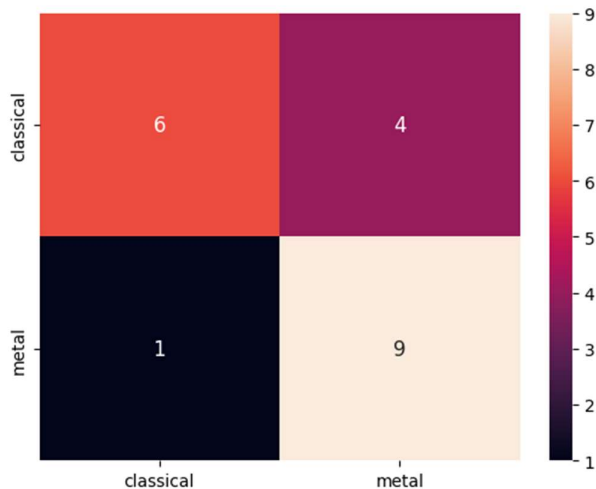

Fig. 4. Confusion matrix for SVC kernel poly

The table below summarizes the accuracy of all the algorithm applied.

|   | algorithm | accuracy |
|---|---|---|
| 0 | LogisticRegression | 0.95 |
| 1 | SVC sigmoid | 0.95 |
| 2 | KNeighborsClassifier 8 | 0.90 |
| 3 | RandomForestClassifier | 0.90 |
| 4 | DecisionTreeClassifier | 0.90 |
| 5 | GaussianNB | 0.90 |
| 6 | SVC linear | 0.90 |
| 7 | SVC rbf | 0.90 |
| 8 | SVC poly | 0.75 |

Fig. 5. Accuracy after feature engg. using LDA

B. *Part 2 Feature Selection*

An expert opinion is needed to select features to run the models. The number of features needs to be less than the number of samples. Here out of 138 we need to have less than 10 features selected. In the analysis we have selected feature 'spectral_centroid_mean' and 'energy_entropy_mean'. In future works better feature selection can be done and more combination of features can be explored. We can plot and see the distribution and get an idea on models to train.

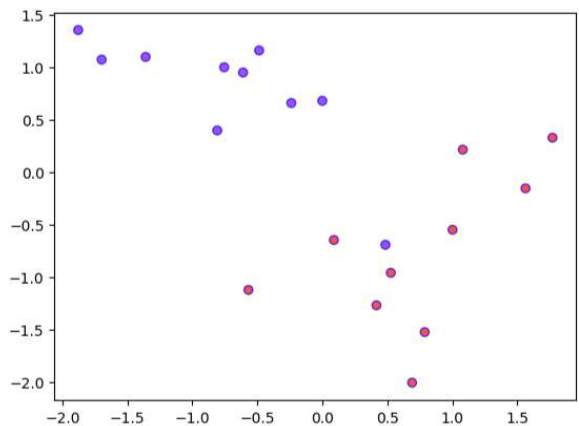

Fig. 6. Feature plot of selected 2 features.

From fig 4 it seems that linear classification models will do good. Next we try to apply to KNN.

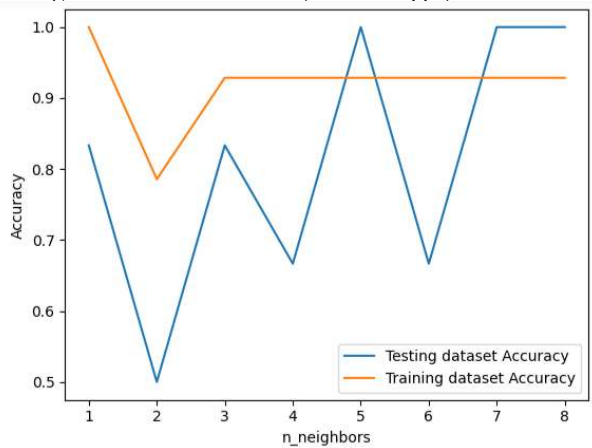

Fig. 7. Value of k vs Accuracy.

From fig 5 above the value of k can be 7 or greater. We have taken k=7 to train the model. For the calculation of accuracy on the trained model we use the complete dataset and not the test only dataset. This due to the less number of samples.

The table below summarizes the accuracy of the algorithms trained.

| | algorithm | accuracy |
|---|---|---|
| 0 | LogisticRegression | 0.95 |
| 1 | SVC sigmoid | 0.95 |
| 2 | KNeighborsClassifier 7 | 0.95 |
| 3 | RandomForestClassifier | 0.95 |
| 4 | DecisionTreeClassifier | 0.90 |
| 5 | GaussianNB | 0.95 |
| 6 | SVC linear | 0.95 |
| 7 | SVC rbf | 0.95 |
| 8 | SVC poly | 0.90 |

Fig. 8. Accuracy after feature selection

From fig 1 and fig 6, data seems to be linearly separable and thus a linear model needs to be trained. This is again proved by fig 5 and fig 8. where logistic regression model comes out be a clean winner.

III. CONCLUSION

Paper shows use ML methods to classify audio into 2 genres, classical and metal. First python library pyAudioAnalysis is used to extract 138 features. Feature engneering and Feature selection has been applied to create a dataset on which ML methods KNN, SVC, Logistic Regression, Random Forest, DecisionTreeClassifier and GaussianNB is applied.

In future study more number of audio files can be taken as opposed to 20 audio files with more than 2 classes of genre to classify. An expert opinion can be taken to do feature selection. As an alternate LDA can also be used for feature engineering as shown.

ACKNOWLEDGMENT

I am thankful to professor Dr. JB Simha for teaching AI ML and Speech Analytics as part of course curriculum of M. Tech in Artificial Intelligence at REVA Academy for Corporate Excellence.